\documentclass[12pt,a4paper,twoside]{paper}
\usepackage{amsmath,bbm,amssymb,amsxtra}
\usepackage{graphicx}
\usepackage{epstopdf}
\usepackage{chngcntr}
\usepackage{etoolbox}

\usepackage{bbold}

\usepackage{fancyhdr}
 \usepackage{fancyhdr}
\newcommand{\be}{\begin{equation}}
\newcommand{\ee}{\end{equation}}

\newcommand{\ba}{\begin{eqnarray}}
\newcommand{\ea}{\end{eqnarray}}

\newcommand{\bea}{\begin{eqnarray*}}
\newcommand{\eea}{\end{eqnarray*}}

\newcommand{\bee}{\begin{enumerate}}
\newcommand{\ene}{\end{enumerate}}

\usepackage{epsfig}
\def\R{\mathbb R}

\usepackage{appendix}
\usepackage{chngcntr}
\usepackage{etoolbox}
\usepackage{lipsum}
\usepackage{amsmath,bbm,amssymb,amsxtra}
\usepackage{enumerate}
\allowdisplaybreaks
\usepackage[active]{srcltx}
\usepackage{epsfig}
\usepackage{ae}
\numberwithin{equation}{section}

\newtheorem{lemma}{Lemma}[section]
\newtheorem{proposition}[lemma]{Proposition}

\newtheorem{theorem}[lemma]{Theorem}

\newtheorem{definition}[lemma]{Definition}

\newcommand{\prop}[1]{\begin{proposition}\label{#1}
\sl }
\newcommand{\eprop}{\end{proposition}}
\newcommand{\thm}[1]{\begin{theorem}\label{#1}
\Ä }
\newcommand{\ethm}{\end{theorem}}
\newcommand{\lem}[1]{\begin{lemma}\label{#1}
\sl }
\newcommand{\elem}{\end{lemma}}

\newcommand{\defin}[1]{\begin{definition}\label{#1}
\sl }
\newcommand{\edefin}{\end{definition}}

\def\la{\lambda}

\def\CA{{\mathcal A}}              
       \def\CE{{\mathcal E}}       
       \def\CH{{\mathcal H}}

\def\CH{{\mathcal H}}

\def\qq{ \begin{eqnarray} }
\def\qqq{ \end{eqnarray} }
\def\rr{ \begin{equation} }
\def\rrr{ \end{equation} }

\def\qq{ \begin{eqnarray} }
\def\qqq{ \end{eqnarray} }

\usepackage{epsfig}

\counterwithin*{equation}{section} 
\counterwithin*{equation}{subsection} 

\begin{document}

    \begin{center}
 \textsl{\huge    Schr\"odinger's  paradox and proofs of nonlocality using only perfect correlations}

 \vspace*{8mm}
 { \Huge Jean Bricmont\footnote{IRMP,
Universit\'e catholique de Louvain,
chemin du Cyclotron 2,
1348 Louvain-la-Neuve,
Belgium. E-mail: jean.bricmont@uclouvain.be},
Sheldon Goldstein\footnote{Department of Mathematics, Rutgers University, Hill Center, 110 Frelinghuysen Road, Piscataway,
NJ 08854-8019, USA. E-mail: oldstein@math.rutgers.edu},
Douglas Hemmick\footnote{66 Boston Drive,
Berlin, MD 21811, USA, E-mail: jsbell.ontarget@gmail.com}
}
\end{center}

\vspace*{10mm}
  { \Large  Dedicated to our mentor, Joel L. Lebowitz, a master of statistical physics  who, when it came to foundational issues of quantum mechanics, was always willing to {\it listen}.}

\begin{abstract}

We discuss proofs of nonlocality based on a generalization by Erwin Schr\"odinger of the argument of Einstein, Podolsky and Rosen. These proofs do not appeal in any way to Bell's inequalities. Indeed, one striking feature of the proofs is that they can be used to establish nonlocality solely on the basis of suitably robust perfect correlations. First we explain that Schr\"odinger's argument shows that locality and the perfect correlations between measurements of observables on spatially separated systems implies the existence of a non-contextual value-map for quantum observables; non-contextual means that the observable has a particular value before its measurement, for any given quantum system, and that any experiment ``measuring this observable" will reveal that value. Then, we establish the impossibility of a non-contextual value-map for quantum observables {\it without invoking any further quantum predictions}. Combining this with Schrödinger's argument implies nonlocality. Finally, we illustrate how Bohmian mechanics is compatible with the impossibility of a non-contextual value-map.

\end{abstract}

\section{Introduction}\label{sec1}

In 1935, three important papers concerning the meaning of quantum mechanics were published. One of them was the EPR paper \cite{EPR}, another was the reply of Bohr to that paper \cite{Bo3}. But there was a third one, by Schr\"odinger, on the ``present situation in quantum mechanics," originally published in German \cite{Sch}. 

Schr\"odinger's paper is usually remembered for his famous ``cat" example (a cat whose quantum state is a superposition of ``cat alive" and ``cat dead"), which Schr\"odinger considered as ``quite ridiculous" and a {\it reductio ad absurdum} of the usual understanding of superposed quantum states, since, for him, the cat was obviously either dead or alive, {\it whether one looks at it or not}.
But, together with two subsequent papers by Schr\"odinger \cite{Sch1, Sch2}, it also contained an important generalization of the argument given by EPR and an extended discussion of its significance.

Almost thirty years later, in 1964, Bell \cite{Be2} derived his now famous inequality which, combined with the EPR argument, proves that there exist nonlocal effects in nature. We are of course aware that not everybody agrees with that conclusion (and that is quite an understatement). One of the goals of this paper is to prove, via another route, the inevitability of that very same conclusion. 
Our aim here is to combine Schr\"odinger's generalization of the EPR argument with some  no hidden variables theorems in order 
to prove the existence of nonlocal effects in nature. This paper is in large part based on the PhD thesis of one of us \cite{Hem, Hem-S}. Other papers using arguments similar to ours  are \cite{St, H-R, Brown, El, Ara, Cab}; we will discuss them in section \ref{sec7}.

We  first recall the EPR argument (Section \ref{sec2}), then explain its generalization by Schr\"odinger (Section \ref{sec3}) and the no hidden variables theorems that we use (Section \ref{sec4}). In particular, we point out that the impossibility of a value-map for quantum observables follows merely from the requirement that the map be non-contextual (Section \ref{sec4}).
From that, nonlocality follows (Section \ref{sec5}), indeed based solely on perfect correlations. Finally, we will explain how both the impossibility of a value-map and nonlocality can be understood within the pilot-wave theory or Bohmian mechanics (Section \ref{sec6}).

\section{The essence of the EPR argument}\label{sec2}

Let us first consider an experimental situation that may, at first sight, seem baffling: one measures, in a given direction, the spin of a  particle   that is  prepared in such a way that, according to quantum mechanics, the result will be ``random" with both outcomes, up and down, equally probable.

Yet, there is someone, call him Bob, who claims to be clairvoyant and who does predict, in each run of the experiment and with 100$\%$ success,  what the result will be.

How does Bob do it? Well, as with every paranormal claim, there is a trick: Bob has an accomplice, Alice, and there is a box between Bob and Alice that sends to each of them one particle belonging to an entangled pair, which means that their joint quantum state $|\Psi\rangle$ is of the form:

\begin{eqnarray}
|\Psi\rangle &=& \frac{1}{\sqrt 2} \big(|  \uparrow  \rangle| \downarrow  \rangle-|  \downarrow  \rangle|   \uparrow  \rangle\big),
\label{1}
\end{eqnarray}
where the left factors refer to the particle sent to Alice and the right to the one sent to Bob.
That state, according to ordinary quantum mechanics, means that the spin measured by Bob will have equal probability to be up or down, but is perfectly anti-correlated with the one of Alice: if the spin is up for Alice, it will be down for Bob and vice-versa.

So the trick is simply that Alice measures the spin of her particle first and sends to Bob  a message giving her result.\footnote{For the reader who might worry that, according to special relativity, the temporal order, ``before" and ``after," depends on the reference frame for spatially separated events, let us say that everything here refers to the reference frame where the laboratory in which the measurements are performed is at rest. See \cite{Ma} for a discussion of the tensions that exist, due to such experiments,  between quantum mechanics and relativity.} Then Bob can, with certainty, predict what the result of the spin measurement on his side will be, namely the opposite of the one of Alice.

So, there is no mystery here and, within ordinary quantum mechanics, this is ``explained" by the collapse rule: when Alice measures the spin of her particle, the state $|\Psi\rangle$ collapses to $|  \uparrow  \rangle|  \downarrow  \rangle$ if Alice sees the spin up and to 
$|  \downarrow  \rangle|   \uparrow  \rangle$ if she sees the spin down.

And, after the collapse, the spin of the particle on Bob's side is no longer undetermined: it is, with probability one, up if the collapsed state is  $|  \downarrow  \rangle|   \uparrow  \rangle$ and down if the collapsed state is  $|  \uparrow  \rangle|  \downarrow  \rangle$.

But  one may ask, what does this collapse mean? Is it a physical operation, i.e. does the measurement by Alice change the physical situation on Bob's side,
or does it have to do only with ``information," meaning that, through the measurement by Alice,  we simply {\it learn} something about the physical situation on Bob's side?

It is important, for the rest of the discussion, to keep that dilemma, which we will call {\it the EPR dilemma}, in mind and, if one chooses a horn of that dilemma, to stick to it and to its meaning. A lot of confusion comes from the fact that some people jump from one horn of the dilemma to the other in the course of a discussion.

Consider the first horn of the dilemma. What could this physical operation be? One may think that, before Alice's measurement, the spin of the particle on Bob's side is genuinely undetermined or non-existent and that it acquires a given value only after Alice's measurement.

But that means that some sort of nonlocal action or action at a distance took place.

According to  the second horn of the dilemma, the spin of the particle on Bob's side was determined before Alice's measurement and we simply {\it learn} its value by doing that measurement. But that also means that quantum mechanics is ``incomplete" in the sense that EPR gave to that term: there are facts of the matter about the spin of the particle on Bob's side (it is either up or down) that are not described by the superposed state (\ref{1}).

There are three differences, one minor and two major ones, between our treatment here of the EPR argument and the original one.

The minor difference is that EPR did not discuss spin but rather position and momentum. The reformulation of their argument in terms of spin is due to David Bohm \cite{Bo}. 

One major difference is that EPR discussed two variables or ``observables" (position and momentum) and so do most people who use the spin variables or ``observables," while we use {\it only one} variable, and that is actually enough to pose what we call the EPR dilemma.\footnote{This argument is discussed in detail by Maudlin \cite[3rd edn, pp.~128--132]{Ma}.}

Another major difference with respect to EPR  is that they did not really pose their dilemma as such, because, for them, one of its horns, involving nonlocality, was unthinkable. They thought  that they had demonstrated the incompleteness of quantum mechanics, but, nevertheless,  the EPR dilemma was implicitly an essential part of their argument.

\section{Schr\"odinger's generalization of the EPR argument}\label{sec3}

The main innovation of Schr\"odinger in \cite{Sch, Sch1, Sch2}, where he coined the expression ``entangled state," was to extend the EPR-type argument of the previous section  to {\it all} observables, at least given suitable quantum states, that we will discuss now.

\subsection{Maximally entangled states}\label{sec3.1}

Let us start by mentioning a theorem that we will not use later but that sets the stage for the definition of maximally entangled states. 

Consider a finite dimensional\footnote{For the proofs of nonlocality, it will be sufficient to consider finite dimensional spaces, which allows us to avoid all mathematical  subtleties.} (complex) Hilbert space $\cal H$, of dimension $N$,  and let $\Psi$ be a unit vector in the tensor product $\cal H \otimes \cal H$.

Then, a special case of the Schmidt decomposition theorem (see e.g. \cite[sect.2.5]{NC}) states that there exist orthonormal bases $\psi_n$ and $\phi_n$ in $\cal H$ (we will assume below that all bases are orthonormal) and non negative real coefficients $c_n$, so that one can write:

\be
\Psi= \sum_{n=1}^N c_n   \psi_n \otimes \phi_n ,
\label{Schm}
\ee
with $\sum_{n=1}^N |c_n|^2=1$.

A unit vector in $\cal H \otimes \cal H$ is {\it maximally entangled} if it is of the form (\ref{Schm}) with $c_n=\frac{1}{\sqrt N}$, $\forall n= 1, \dots, N$:
\be
\Psi= \frac{1}{\sqrt N}\sum_{n=1}^N    \psi_n \otimes \phi_n.
\label{ME}
\ee

 Since we are interested in quantum mechanics, we will refer to those vectors as {\it maximally entangled states} and we will associate, by convention, each space in the tensor product to a ``physical system," namely we will consider the set  $\{\phi_n\}_{n=1}^N$ as a basis of states for physical system 1 (associated to Bob when measurements are made on that system) and the set  $\{\psi_n\}_{n=1}^N$ as a basis of states for physical system 2 (associated to Alice when measurements are made on that system).

Now, given a  maximally entangled state, we  will  associate to each operator of the form ${\mathbb 1}  \otimes O$ (meaning that it acts non-trivially only on particle 1) an operator of the form $ \tilde O \otimes  {\mathbb 1}  $ (meaning that it acts non-trivially only on particle 2). Here ${\mathbb 1}$ denotes the identity operator on  $\cal H$.

First, define the operator $U$ mapping $\cal H$ to $\cal H$ by setting
\be
U \phi_n =\psi_n,
\label{defU}
\ee
$\forall n= 1, \dots, N$, and extending $U$ to an anti-linear operator on all of $\cal H$:
\be
U (\sum_{n=1}^N c_n \phi_n) =\sum_{n=1}^N c^*_n U \phi_n= \sum_{n=1}^N c^*_n \psi_n
\label{U1}
\ee
where $^*$ denotes the complex conjugate.

Then U is an anti-unitary operator, meaning that it is anti-linear and satisfies, $\forall  \psi,  \phi \in \CH$:
\be
\langle U \psi |  U \phi \rangle = \langle  \psi |   \phi \rangle^*=  \langle  \phi |   \psi \rangle.
\label{U2}
\ee

A simple example of an anti-unitary operator is given by the (basis dependent)  complex conjugation $C$:
\be
C \phi= \phi^*
\label{Co}
\ee
In fact, every anti-unitary operator can be written as $U= C \tilde U$, where $\tilde U$ is unitary (for any choice of complex conjugation C).

Using the operator $U$, the state $\Psi$ in (\ref{ME}) can be written as:
\be
\Psi= \frac{1}{\sqrt N}\sum_{n=1}^N   U\phi_n \otimes \phi_n .
\label{ME1}
\ee
It is easy to check that this formula is the same for any basis. Let $\{\chi_n\}_{n=1}^N$ be another basis of $\cal H$.
We can write: $\phi_n= \sum_{k=1}^N \langle \chi_k | \phi_n  \rangle\chi_k$ and, inserting this in (\ref{ME1}), we obtain that
\ba
\nonumber
\Psi &=& \frac{1}{\sqrt N}\sum_{n=1}^N   U(\sum_{k=1}^N \langle \chi_k  | \phi_n \rangle\chi_k) \otimes \sum_{k'=1}^N \langle \chi_{k'}    | \phi_n   \rangle\chi_{k'}\\\nonumber
&=& \frac{1}{\sqrt N}\sum_{n=1}^N   \sum_{k=1}^N \langle  \phi_n  | \chi_k \rangle U \chi_k \otimes \sum_{k'=1}^N \langle\chi_{k'} | \phi_n  \rangle\chi_{k'}\\\nonumber
&=& \frac{1}{\sqrt N}   \sum_{k=1}^N  \sum_{k'=1}^N \langle \chi_{k'}  \sum_{n=1}^N  |\phi_n \rangle \langle \phi_n | \chi_{k} \rangle U \chi_k  \otimes \chi_{k'}\\\nonumber
&=& \frac{1}{\sqrt N}   \sum_{k=1}^N  \sum_{k'=1}^N \langle \chi_{k'}   | \chi_{k} \rangle U \chi_k  \otimes  \chi_{k'}\\
&=& \frac{1}{\sqrt N}   \sum_{k=1}^N   U \chi_k  \otimes  \chi_{k},
\label{ME2}
\ea
where in the second equality we used the anti-linearity of $U$, in the fourth equality we used that $\sum_{n=1}^N |\phi_n \rangle   \langle \phi_n|= \mathbb{1}$ and in the last one we used that
 $\langle \chi_k   | \chi_{k'} \rangle=\delta_{k, k'}$.
 
$U$ thus determines, and is uniquely determined by, a maximally entangled state $\Psi$.

Given a maximally entangled state $\Psi$, and hence $U$, we may associate
 to every operator of the form  ${\mathbb 1}  \otimes O$ an operator of the form  $ \tilde O \otimes {\mathbb 1}$ by setting
\be
\tilde O = U O  U^{-1}.
\label{A}
\ee

If $\phi_n$ are eigenstates of $O$, with eigenvalues $\la_n$,
\be
 O \phi_n = \la_n \phi_n.
\label{A1}
\ee
Then, the states $\psi_n= U \phi_n$ are eigenstates of $\tilde O$, also with eigenvalues $\la_n$:
\be
\tilde O \psi_n = \la_n \psi_n.
\label{A2}
\ee
This implies and is in fact equivalent to the following relationship between the operators $O$ and $\tilde O$:
\be
( O \otimes  {\mathbb 1} - {\mathbb 1} \otimes \tilde O ) \Psi = 0,
\label{A3}
\ee
directly expressing the fact that the state $ \Psi$, $O \otimes  {\mathbb 1}$ and ${\mathbb 1} \otimes \tilde O$
are perfectly correlated.

We have proven the following
\begin{theorem}\label{0}
Consider a finite dimensional Hilbert space $\cal H$, of dimension $N$,  and a  maximally entangled state $\Psi \in \cal H \otimes \cal H$. Then, for any
self-adjoint operator $O$ acting on $\cal H$,
 there exists a self-adjoint operator  $\tilde O$ acting on $\cal H$ such that (\ref{A3}) holds.
\end{theorem}

{\bf Remarks}

\begin{itemize}
\item[1.] 

If $O$ has eigenstates $\phi_k$ and the maximally entangled state (\ref{ME}) has $\psi_n= \phi_n$, then $\tilde O=O$.
In general $\tilde O$ will of course be different from $O$. Here is a simple example in which, while different, they are almost the same: 
\be
\nonumber
\phi_1= | \uparrow>,\;\; \phi_2 = | \downarrow>, \;\;\psi_1= -|\downarrow>, \;\; \psi_2=|\uparrow>. 
\ee
One obtains that
\ba
\nonumber U |  \uparrow>&=& -|  \downarrow>,\\\nonumber
U |  \downarrow>&=& |  \uparrow>.
\ea
so that $U= C \tilde U$, where $C$ is the complex conjugation in the $\phi$ basis and 
\ba
\tilde U = \left(\begin{array}{ccc} 0 & 1 \\ -1 & 0 \end{array}\right).
\ea
If one takes 
\ba
O = \left(\begin{array}{ccc} 1 & 0 \\ 0 & -1 \end{array}\right)
\ea
which has eigenvectors $\phi_1$ with eigenvalue $1$ and $\phi_2$ with eigenvalue $-1$, since
\ba
{\tilde U}^{-1} = \left(\begin{array}{ccc} 0 & -1 \\ 1 & 0 \end{array}\right),
\ea
one  finds that
\ba
\tilde O =  U O  U^{-1}= \tilde U O {\tilde U}^{-1}= \left(\begin{array}{ccc} -1 & 0 \\ 0 & 1 \end{array}\right)=-O.
\ea

We will use later the following  

\item[2.] 

{\it Products of maximally entangled states are maximally entangled states}: If one has two Hilbert spaces $ {\cal H}_1$, $ {\cal H}_2$, and two   maximally entangled states $\Psi_i \in {\cal H}_i \otimes {\cal H}_i$, $i= 1, 2$, then it is easy to check that the state $\Psi= \Psi_1 \otimes \Psi_2$ is maximally entangled  in ${\cal H} \otimes {\cal H}$, where
${\cal H}= {\cal H}_1 \otimes {\cal H}_2$ (under the canonical identification of $({\cal H}_1 \otimes {\cal H}_2) 
\otimes ({\cal H}_1 \otimes {\cal H}_2)$ with $ {\cal H}\otimes {\cal H}$).

\end{itemize}

\subsection{Schr\"odinger's theorem}\label{sec3.2}

Let us now generalize the EPR argument  by applying the result of the previous subsection to spatially separated physical systems.

Suppose that we have a pair of physical systems, whose states belong to the same finite dimensional  Hilbert space $\cal H$. And suppose that the quantum state $\Psi$ of the pair is maximally entangled, i.e. of the form (\ref{ME}).

Any ``observable" acting on system 1 is represented by a self-adjoint operator $O$, which has therefore a basis of eigenvectors. Since the representation (\ref{ME1}) of the state $\Psi$ 
is valid in any basis, we may choose, without loss of generality, as the set $\{\phi_n\}_{n=1}^N$  in (\ref{ME})  the eigenstates of $O$. Let  $\la_n$  be the corresponding  eigenvalues, see  (\ref{A1}). 

 If one measures that observable $O$, the result will be one of the eigenvalues $\la_n$, each having equal probability $\frac{1}{N}$. If the result is $\la_k$,  the (collapsed) state of the system after the measurement, will be $\psi_k \otimes \phi_k $. Then, the measurement of observable $\tilde O$, defined by (\ref{A}, \ref{defU}), on system 2, will necessarily yield the value $\la_k$. 
 
Reciprocally, if one measures an observable $\tilde O$ on system 2 and the result is  $\la_l$,  the (collapsed) state of the system after the measurement, will be $\psi_l \otimes \psi_l $, and the measurement of observable $ O$ on system ~1 will necessarily yield the value $\la_l$. 

Let us summarize what we just said:

{\bf Principle of perfect correlations.}
{\it In any maximally entangled  quantum state, of the form (\ref{ME}), there is, for each operator $O$ acting on system 1, an operator $ \tilde O$ acting on system 2 (defined by (\ref{A}, \ref{defU})), such that, if one measures the physical quantity represented by operator $ \tilde O$ on system 2 and the result is the eigenvalue  $\la_k$ of $ \tilde O$, then, measuring the physical quantity represented by operator 
$O$  on system 1 will yield with certainty the same eigenvalue $\la_k$, and vice-versa.\footnote{The correlations mentioned here are often called  anti-correlations, for example when $\tilde O=-O$, as in the example of the spin in section \ref{sec2}.}}

The following property  will be crucial in the rest of the paper.
 
{\bf Locality.}
{\it If systems 1 and 2 are spatially separated from each other, then measuring an observable on system 1 has no instantaneous effect whatsoever on system 2 and  measuring an observable on system 2 has no instantaneous effect whatsoever on system 1.}

Finally, we must also define: 
 
 {\bf Non-contextual value-maps.}
 Let $\cal H$ be  a finite dimensional  Hilbert space  and let $\CA$ be the set of self-adjoint operators on $\CH$. Suppose $\cal H$ is the quantum state space for a physical system and $\CA$ is the set of quantum observables. Suppose there are situations in which, like the spin components in the EPR situation, there are observables $A$ for which the result of measuring $A$ is determined already, before the measurement. Suppose, that is, that $A$ has, in these situations, a pre-existing value $v(A)$ revealed by measurement and not merely created by measurement. Of course, this implies that for every experiment ${\cal E}_A$ measuring $A$, the result $v({\cal E}_A)$ of that experiment, in the situation under consideration, must be $v(A)$. And suppose finally that the situation is such that we have such a pre-existing value $v(A)$ for every $A\in \CA$. 
 
 We would then have a 
 {\it  non-contextual value-map}, namely  a map  $v: \CA \to \R$ that assigns the value $v(A)$ to  any experiment associated with what is called in quantum mechanics a ``measurement of an observable A." There can be different ways to measure the same observable. The  value-map is called non-contextual  because all such experiments, associated with the same quantum observable A, are assigned the same value.

This notion of non-contextual value-map is not a purely mathematical one, since it involves the notion of an experiment that measures a quantum observable $A$, which we have not mathematically formalized. However, we shall need only the following obvious purely mathematical consequence of non-contextuality.

 A non-contextual value-map has the following fundamental property: Suppose that, if $A_i$, $i=1, \dots, n$, are mutually commuting self-adjoint operators on $\CH$, $
 [A_i, A_j]= 0$, $\forall i, j=1, \dots, n$,
 and that  $f$ is a function of $n$ variables. Then  if $B= f(A_1, \dots, A_n)$, we also have that 
 \be
v(B)= f(v(A_1), \dots, v(A_n)).
\label{res}
\ee

It is a well-known property of quantum mechanics that, since all the operators $A_1, \dots, A_n, B$ commute, they are simultaneously measurable and the result of those measurements must satisfy (\ref{res}).

But, and this is what we want to emphasize, (\ref{res}) follows trivially from the non-contextualilty of the value-map. Indeed, a valid quantum mechanical  way to measure the operator $B= f(A_1, \dots, A_n)$ is to measure $A_1, \dots, A_n$ and, denoting the results $\la_1, \dots, \la_n$, to regard $\la_B=f(\la_1, \dots, \la_n)$ as the result of a measurement of $B$ .
Since, by the non-contextuality of the map $v$, all the possible measurements of $B$ must yield the same results, (\ref{res}) holds.

Thus, once one has a non-contextual value-map, which, as we shall discuss, is a consequence of the perfect correlations and locality, 
 {\it one does not even need to check (\ref{res})}.

Now we will use the perfect correlations and locality to establish the existence of a non-contextual value-map $v$,
for a maximally entangled  quantum state of the form (\ref{ME}) or, equivalently, (\ref{ME1}). By the 
principle of perfect correlations,
for any  operator $O$ on system 1, there is an  operator  $\tilde O$  on system 2, defined by (\ref{A}, \ref{defU}), with which it is perfectly correlated to $O$ through (\ref{A3}).

Thus if we were to measure  $\tilde O$, obtaining  $\la_l$, we would know that 
\ba
v(O)= \la_l. 
\label{map}
\ea
concerning the result of then measuring $O$. Therefore $v(O)$ would pre-exist the measurement of $O$.
But, by the assumption of locality, the measurement of   $\tilde O$, associated to the second system,  could not have 
had any effect on the first system,
and thus, this value $v(O)$ would pre-exist also  the measurement of  $\tilde O$ and this would not depend upon whether $\tilde O$ 
had been measured.
Therefore  the map $O\to v(O)$ where $O$ ranges over all operators on system 1,  
is a non-contextual value-map. 

To summarize, we have shown:

\noindent
{\bf Schr\"odinger's ``Theorem".} Let $\CA$ be the set of self-
adjoint operators on the component Hilbert space   $\CH$ of a physical system 
in a maximally entangled state (\ref{ME}). Then, assuming locality and the principle of perfect correlations, there 
exists a non-contextual value-map  $v: \CA \to \R$.

{\bf Remarks}

\begin{itemize}
\item[1.] 
We put ``Theorem" in quotation marks because the  statement concerns physics and not just mathematics. Its conclusions are nevertheless inescapable assuming the validity of locality and the principle of perfect correlations, which is implied by the quantum formalism. 
 When Schr\"odinger derived this ``theorem" \cite{Sch} he was deeply puzzled by it, and that---in addition to the fact that it is rather common to refer to the EPR paradox---is why we call it Schr\"odinger's paradox, although he did not use that expression. Like EPR, Schr\"odinger took locality for granted, so that he thought that he had established  an  incompleteness of quantum mechanics even more radical than what EPR had claimed to have found.
\item[2.] 
In the literature on quantum mechanics, the values taken by the map $v$ are often called ``hidden variables," because they complement the description of a  physical system given by its quantum state. But we prefer the term ``value-map" because, as we will discuss in Section \ref{sec6}, the expression ``hidden variables" is really a misnomer. 
\item[3.] 
The existence of this map leads naturally to what one may call the naive  statistical interpretation of quantum mechanics. The meaning of the state (\ref{ME}) would be, according to this interpretation, that, for an ensemble of systems each of which has (\ref{ME}) as its quantum  state, the statistical distribution of $v(O)$ would be given by 
$\frac{1}{N}$ for each possible value $\la_k$, $k=1, \dots, N$, of $v(O)$. More generally, for a state like  (\ref{Schm}), it would be given by $|c_k|^2$, which is the Born rule.
\item[4.] 
A frequent objection to our ``theorem" is that, given two non-commuting observables, $\tilde O$ and $ \tilde O'$,  one can measure only one of them at a time and that the measurement of $\tilde O$ will in general affect the result of the measurement  of $\tilde O'$. Hence, goes the objection, one cannot attribute a value $v(O)$ to all observables at once.\footnote{This is one way to understand Bohr's response to EPR \cite{Bo3}, although that response was not very clear.} But that objection ignores the locality assumption which implies that  the choice of the experimentalist acting on  system 2 cannot possibly affect the properties of system 1. 

Schr\"odinger illustrated this with the following analogy \cite{Sch}: suppose that we have a large set of schoolchildren to which one of two questions can be asked (the questions being similar to the choice between two operators $\tilde O$ and $ \tilde O'$), the questions being chosen at random (say, by coin tossing) and suppose that the children always give the correct answer to the question being asked of them.\footnote{The correct answer here being the value $v(O)$ or $v(\tilde O')$ which is determined by the measurement of $\tilde O$ or of $ \tilde O'$.} Would anybody doubt that this means that the children {\it know the answers to both questions that they could possibly be asked}? Would anybody suggest that the question being selected at random is precisely the one that they know the answer to and that, if the other question had been selected, they would not have been able to answer? Of course this could happen ``by accident." Students passing an exam can be lucky and be given just the questions for which they know the answers and not any other ones. But if that experiment is repeated a very large number of times, the possibility of an ``accident" becomes less and less likely. And here, quantum mechanics predicts that the correct answer (the value $v(O)$ or $v( \tilde O')$ determined by the measurement of $\tilde O$ or of $ \tilde O'$)
 will be given no matter how many times the experiment is repeated.

 \end{itemize}

\section{The non-existence of non-contextual value-maps}\label{sec4}

The problem posed by the non-contextual value-map $v$ whose existence is implied by Schr\"odinger's ``theorem" is that such maps simply do not exist (and that is a purely mathematical result). Indeed, one has the:

\noindent
{\bf ``Theorem": Non-existence of non-contextual value-maps.}
Let $\CA$ be the set of self-adjoint operators on the  Hilbert space $\cal H$ of a physical system. Then there exists no non-contextual value-map $v: \CA \to \R$.
 
This ``theorem" is an immediate consequence of the following theorem, since (\ref{res2}, \ref{res3}) are consequences of (\ref{res}).\footnote{This is obvious for (\ref{res3}),  a special case of (\ref{res}).
 For (\ref{res2}) we observe that, since $O$
is self adjoint, we can write 
$O = \sum_i \lambda_i P_{\lambda_i}$ where $P_{\lambda_i}$ is the projector on the subspace
of eigenvectors   of eigenvalue $\lambda_i$ of $O$
and thus we have that $f(O) = \sum_i f(\lambda_i) P_{\lambda_i}$. If we choose any $f$ whose range is the set of  eigenvalues of $O$ and is such that $f(\lambda_i)= \lambda_i$ $\forall i$, we have that 
$O=f(O)$ and, 
by (\ref{res}), we obtain that $v(O)=v(f(O))=f(v(O))$ and thus  $v(O)$ is an eigenvalue of $O$.}

\begin{theorem}\label{2} Let $\cal H$ be  a   Hilbert space of dimension at least three,  and let $\CA$ be the set of self-adjoint operators on $\CH$. There does not  exist a map  $v: \CA \to \R$ such that:

1) $ \; \forall O \in { \CA}$, 
\be
v(O) \;\;\mbox{is an eigenvalue of} \;\;O.
\label{res2}
\ee

2) $\forall O, O' \in { \CA}$ with  $ [O, O']= OO'-O'O=0$, and
for any real valued function $f$ of two real variables,
\be
v(f(O, O'))=f(v(O), v(O')).\label{res3}\\
\ee
\end{theorem}

For the non-existence of a value-map satisfying (\ref{res3}) it is enough to choose for $f$ either
$f(x, y)=xy$ or $f(x, y)= x+y$. Thus theorem \ref{2} is a consequence of either  of the following stronger  non-existence results.\footnote{This is not quite true. For the choice $f(x, y)=xy$ we must have $\dim \CH \geq 4$.}
\begin{theorem}\label{2a}
 Let $\cal H$ be  a   Hilbert space of dimension at least four,  and let $\CA$ be the set of self-adjoint operators on $\CH$.
There does not  exist a map $v: \CA \to \R$ such that:

$1) \; \forall O \in { \CA}$, 
\be
v(O) \;\;\mbox{is an eigenvalue of} \;\;O. \nonumber
\ee

2) $\forall O, O' \in { \CA}$ with  $ [O, O']= OO'-O'O=0$, 
\be
v(O O'))=v(O) v(O').\label{res3a}\\
\ee
\end{theorem}

\begin{theorem}\label{2b}
 Let $\cal H$ be  a  Hilbert space of dimension at least three,  and let $\CA$ be the set of self-adjoint operators on $\CH$.
There does not  exist a map $v: \CA \to \R$ such that:

$1) \; \forall O \in { \CA}$, 
\be
v(O) \;\;\mbox{is an eigenvalue of} \;\;O \nonumber
\ee

2) $\forall O, O' \in { \CA}$ with  $ [O, O']= OO'-O'O=0$, 
\be
v(O +O')=v(O) + v(O'),\label{res3b}\\
\ee
\end{theorem}

For a simple proof of  theorem \ref{2a} due to David Mermin, see \cite{Me4}.

{\bf Remark}

 Theorem \ref{2b} is due to John Bell \cite{Be1} and to Kochen and Specker \cite{KS}.
To prove the theorem, we do not need to assume that the map is defined on all operators in $\CA$. For example, the Kochen and Specker proof \cite{KS} uses the squares of spin matrices, $S_x, S_y, S_z$, for spin associated to any  three dimensional set of orthogonal vectors $x, y, z$ in $\R^3$. They have the following properties:
\begin{itemize}
\item[1.]The eigenvalues of $S_x^2$, $S_y^2$ and $S_z^2$ are  $0$  and $1$.
\item[2.] 
$[S_x^2, S_y^2]= [S_y^2, S_z^2]= [S_z^2, S_x^2]= 0.$
\item[3.] 
$
S_x^2+S_y^2+S_z^2=2.$
\end{itemize}

From that and assumptions (\ref{res2}) (\ref{res3b}), it follows that the triple $(v(S_x^2), v(S_y^2), v(S_z^2))$ must be either $(1, 1, 0)$ or $(1, 0, 1)$ or $(0, 1, 1)$.

But that must hold {\it for every}  set of  three dimensional orthogonal vectors $x, y, z$ in $\R^3$. Kochen and Specker were able to exhibit a finite number of such sets so that the above assumption on the values taken by $(v(S_x^2), v(S_y^2), v(S_z^2))$ leads to a contradiction.\footnote{In their original argument, Kochen and Specker used 117 such sets \cite{KS}, but that number was reduced to 33 by Peres \cite{Per, Per1}.}

\section{Nonlocality}\label{sec5}

The conclusions of Schr\"odinger's ``theorem" and of the ``Theorem"  on the non-existence of non-contextual value-maps plainly contradict each other.
So, the assumptions of at least one of them must be false.  Moreover, the stronger Theorem \ref{2} is a purely mathematical result. 
To derive Schr\"odinger's ``theorem," we assume only the perfect correlations and locality. The  perfect correlations are an immediate consequence of quantum mechanics.   The only remaining assumption  is locality. Hence we can deduce:

\noindent
{\bf Nonlocality ``Theorem"}. The locality assumption is false.

We would now like to  clarify our own nonlocality proof by briefly comparing it to the original proof of Bell and
also to the most closely-related results in the literature. We will do so by  briefly discussing  three different ways to establish nonlocality.

All begin with an EPR-type argument establishing, as a consequence of locality and perfect correlations, the existence of pre-existing values of certain quantum observables for a suitable correlated pair of quantum systems, system 1 and system 2. One then shows that the existence of these pre-existing values leads to a contradiction. In the most familiar (Bell-type) proofs of non-locality, one shows that the quantum predictions for the correlations between certain observables for system 1 and certain observables for system 2 conflict with the existence of pre-existing values for these observables, for example   because the correlations for the pre-existing values would have to obey Bell inequalities  incompatible with the quantum correlations.\footnote{ See \cite{Be2} for the original inequalities and  \cite{GNTZ} for a detailed discussion of them. For a simple version of those inequalities see \cite{DGTZ} or 
\cite[Chap. 4]{Bri1}.} Nonlocality would then be established by the experimental verification of the quantum correlations (together with the prior verification of the perfect correlations). 

More recently, beginning with Kochen \cite{K} and most famously associated with the Free Will Theorem of Conway and Kochen \cite{CK} (see section \ref{sec7}), it has been observed that the existence of pre-existing values for certain observables of system 1  alone can lead to a contradiction with the quantum predictions for the results of measuring those observables, for example  the quantum relations between the squares of spin components considered by Kochen and Specker \cite{KS} (see the end of the previous section). Nonlocality would then be established  \cite{Hem, Hem-S, Cab, Ara} by the experimental verification of those  relations (together with the prior verification of the perfect correlations). 

In the nonlocality argument presented here and in \cite{Hem}, we have noted that once the existence of pre-existing values for certain observables of system 1 has been established, one requires no further input from quantum mechanics, and no further experimental verification, to arrive at a contradiction and the conclusion of nonlocality. One merely needs to note that the existence of pre-existing values would constitute a non-contextual value-map, something that is impossible.

Thus we need only the verification of the perfect correlations  predicted by quantum mechanics between  results of measurements  of pairs of observables $O$ and $\tilde O$ associated to any maximally entangled state (see subsection \ref{sec3.1}) to arrive at the conclusion of nonlocality.


We stress, however, that in order to arrive in this way at the 
conclusion of non-locality we must verify that the perfect correlations are sufficiently 
robust: for any
quantum observable $O$ relevant to an impossibility argument such as that of Kochen
and Specker, the perfect correlations between systems 1 and 2 must hold 
for all of the
different experimental procedures for measuring $O$---all of the different 
contexts---that
are implicit in the impossibility proof. (No such robustness is required 
for the usual Bell nonlocality argument.  That argument concerns a {\it single} local measurement procedure for each spin component. The 
conflict with the existence of a value map then arises from the fact 
that the quantum correlations violate Bell's inequality.)

\section{What happens in Bohmian mechanics?}\label{sec6}

In Bohmian mechanics or pilot-wave theory, the complete state of a closed physical system composed of $N$ particles is a pair 
$(|$quantum state$>$, $\bf X)$, where $|$quantum state$>$ is the usual quantum state (given by the tensor product of wave functions with some possible internal states), and 
${\bf X}= (X_1,\ldots, X_N)$ represents the  positions of the particles that exist, independently of whether one   ``looks" at them or one  ``measures" them (each $X_i\in \R^3$).\footnote{ For elementary introductions to that theory, see \cite{Bri2, Tu} and for more advanced ones, see \cite{B, Bo1, BH, Bri1, DGZ, DT,  DGZ1, Go, No}. There are also pedagogical videos made by students in Munich, available at: https://cast.itunes.uni-muenchen.de/vod/playlists/URqb5J7RBr.html.}

These positions are the  ``hidden variables" of the theory, in the sense that they are not included in the purely quantum description $ |$quantum state$>$, but they are not at all hidden: it is only the particles' positions that one detects directly, in any experiment (think, for example, of the impacts on the screen in the two-slit experiment). So, the expression ``hidden variables" is really a misnomer, at least in the context of Bohmian mechanics.

Both objects, the quantum state and the particles' positions, evolve according to deterministic laws, the quantum state guiding the motion of the particles. Thus, since Bohmian mechanics is deterministic, the result of any quantum measurement will be determined beforehand by the quantum state and the configuration of the relevant system. Therefore for Bohmian mechanics there should be a value-map.
Moreover, the experimental predictions of Bohmian mechanics are the same as those of orthodox quantum mechanics. One might thus wonder how this can be compatible with theorem \ref{2}. But the value-map yielded by Bohmian mechanics is contextual, so that theorem \ref{2} does not apply. We illustrate this contextuality with the example of spin.

In Bohmian mechanics, spin is not real. What we mean by this is that, unlike position, corresponding to which the wave function in Bohmian mechanics is supplemented with determinate values $X_i$, no such supplementation is made to deal with spin. Instead, as usual, the wave function gets additional components. 

Nonetheless, in Bohmian mechanics, as in standard quantum mechanics, there are ``spin measurements." Since the theory is deterministic, one might think that, if the initial conditions of the particle whose spin is measured were given (both its quantum state and its initial position), the result of a spin measurement would be determined and  would  define a ``value map" for the spin operator being measured. That is true, but the value map would be contextual, as we will now explain.

From our definition of non-contextual value-maps
in subsection \ref{sec3.2}, we see that it is sufficient, in order to establish the contextuality of the value-map defined here, to find two experiments measuring the same spin operator yielding different  results, for the same initial conditions of the particle (for more details, see \cite[chap. 7]{Al}, \cite[sect. 5.1.4]{Bri1}, \cite{NRAO}).

\begin{figure}[!ht]
\centering
\includegraphics[width=.9\textwidth]{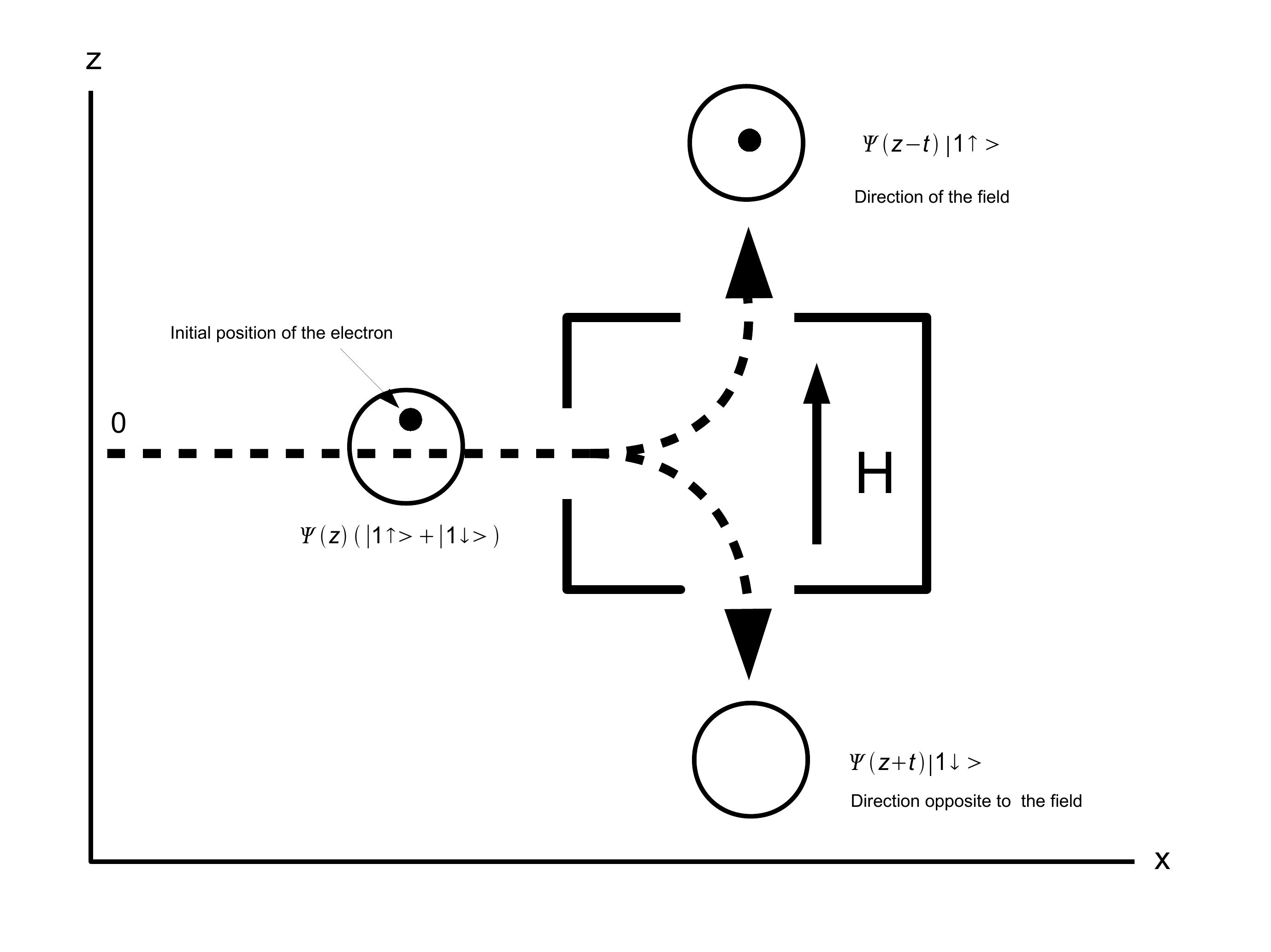}
\caption[]{An idealized spin measurement}\label{3fig4}
\end{figure}
So, consider  a quantum state $\Phi$ which is a superposition of eigenstates
$|1 \uparrow>$ and $|1 \downarrow>$ of $\sigma_z$, the $z$ component of spin, that we describe in an idealized form:  
\ba
\Phi= \Psi(z) (|1 \uparrow> +|1 \downarrow>)
\ea 
$z$ being the vertical direction  (see Figure \ref{3fig4}). We will assume that the spatial part of the state, namely the wave function  $\Psi(z)$, is symmetrical: $\Psi(z)= \Psi(-z)$. The Bohmian dynamics implies then that the particles cannot cross the line  $z=0$. $\Psi$ is also a function of the horizontal variable $x$ (the particle  moves rightwards in the $x$ direction), but we suppress that variable, and we ignore the variable $y$ entirely.

Consider a first procedure $\CE_{\sigma_z}$ to measure the spin in the $z$  direction.
In Figure \ref{3fig4}, $H$ denotes an inhomogeneous magnetic field whose gradient is oriented in that direction; the disks represent (in a very idealized way) the support of the spatial part of the wave function. The   $|1 \uparrow>$ part of the quantum state always goes in the direction of the field (which gives rise to the state $\Psi(z-t) |1 \uparrow>$)
and the  $|1 \downarrow>$  part always goes in the direction opposite to the field  ($\Psi(z+t) |1 \downarrow>$). But the particle, if it starts initially in the part of the wave function above the line $z=0$ (that the particles cannot cross), will always go up. When it does so, we say  that the spin has been measured to be up, i.e. that $\sigma_z=1$.

\begin{figure}[!ht]
\centering
\includegraphics[width=.9\textwidth]{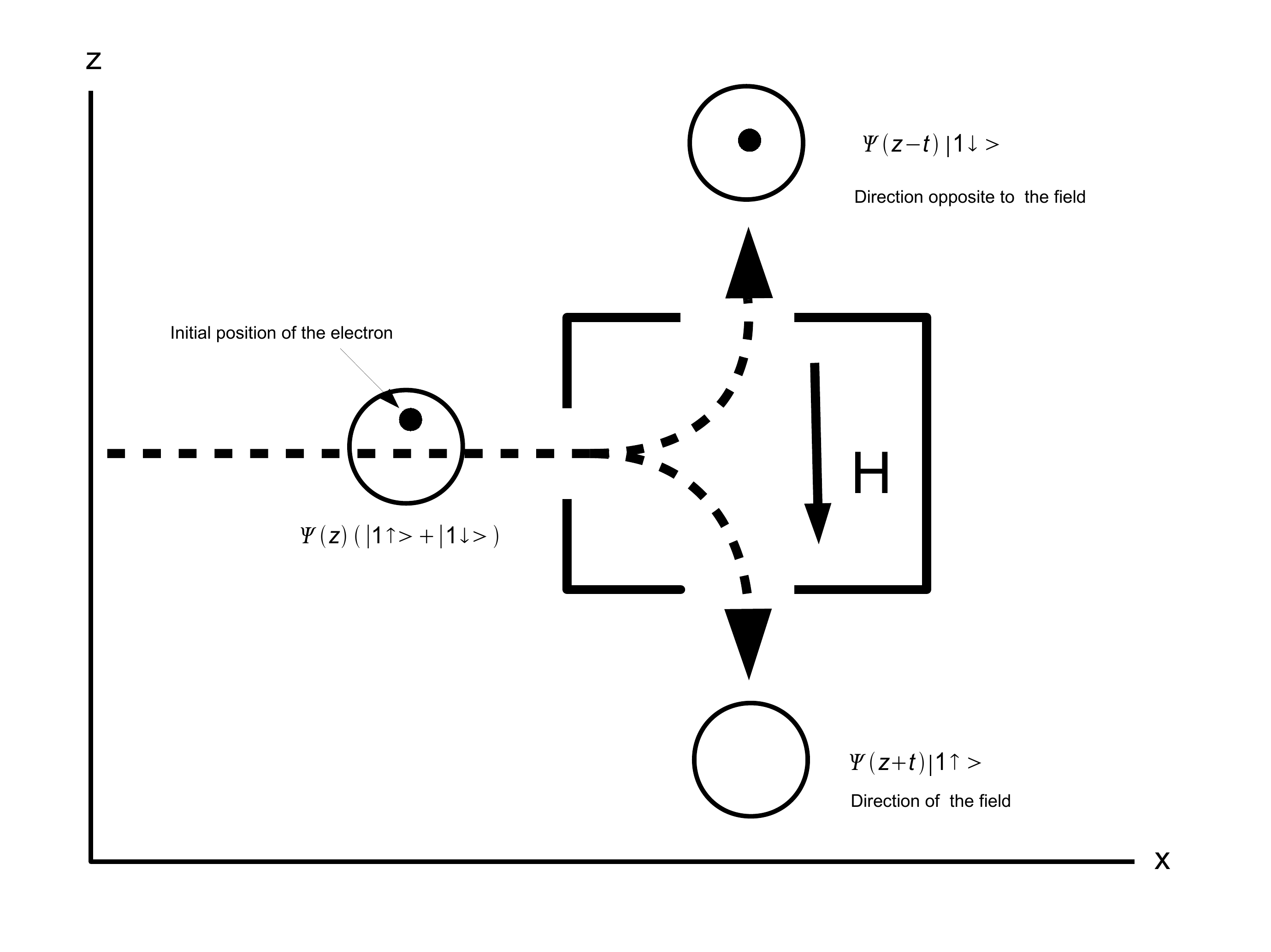}
\caption[]{An idealized spin measurement with the direction of the gradient of the magnetic field reversed with respect to the one of Figure~\ref{3fig4}}\label{3fig5}
\end{figure}  

Consider a second  procedure $\CE'_{\sigma_z}$  to measure the spin in the $z$ direction, namely by reversing the direction  of the  gradient  of the field, for example by reversing the orientation of $H$ while leaving its inhomogeneity otherwise unchanged,
as in Figure \ref{3fig5}. With this change, the spin eigenstates reverse their motion in the $z$ direction, as depicted in  Figure \ref{3fig5}.
This procedure remains a measurement of $\sigma_z$, but with reversed calibration: now we say that
$\sigma_z=-1$ when the particle goes up.

However, since the
 particle cannot cross the  line $z=0$, starting with exactly the same initial position as in  Figure \ref{3fig4}, it will again  go upwards, so that the second procedure yields the value $\sigma_z=-1$ in this case. 
 We have in fact that $v(\CE'_{\sigma_z})=- v(\CE_{\sigma_z})$ with $v(\sigma_z)$ itself undefined. The value map is contextual.

 One can also show that additional contextuality arises from the nonlocal character of Bohmian mechanics (which it must have, since if the theory were local, there would be a non-contextual value map, as we explained in subsection \ref{sec3.2}).
  
 Suppose that one sends  two particles, $A$ and $B$, whose joint quantum state $\Psi$ is of the form (\ref{1}),  towards measuring devices, located far apart, and that one measures the $z$ component of the spin of particle $A$ first, and then that of  particle $B$.  The choice of the way to measure  $\sigma_z$ of particle $A$ (as in Figure \ref{3fig4} or as in Figure \ref{3fig5}) will affect the result of the measurement of $\sigma_z$ for the $B$ particle. As  a matter of fact, assuming sufficient symmetry, the two procedures will yield opposite results for $\sigma_z$ of particle $A$, just as for the case of a single particle. But then, since the particles are perfectly anti-correlated, the results for $\sigma_z$ of particle $B$ will be either $+1$ or $-1$, depending only on which procedure was performed on particle $A$. This is one illustration of how nonlocality manifests itself in Bohmian mechanics.

\section{Conclusions}\label{sec7}

In the literature on Bell's theorem, one often finds the assertion that this theorem forces us to give up either realism or
locality. Realism, viewed as a philosophical attitude, can have several different meanings, the weakest one being the rejection of solipsism (``there is nothing outside my consciousness") or of radical skepticism (``there is a reality but it is unknowable"). In that sense, realists contend that there is a world ``out there" having objective properties and that the goal of science is to discover those properties. If that is what realism means, most physicists are realists: they believe in atoms, stars, galaxies etc. They also believe that measuring devices indicate definite results at the end of an experiment; otherwise, why believe in quantum mechanics in the first place, since all the evidence for quantum mechanics comes from those results?\footnote{One might object that adherents of the ``Many-Worlds Interpretation" of quantum mechanics \cite{Eve, DWG} do not share that view, since, for them, experiments do not have a single outcome. But they nevertheless agree that measuring devices indicate definite results in each World.}

So, the meaning of realism in discussions of Bell's theorem must be different from those generalities and it often refers to what one might call ``microscopic realism," the idea that quantum objects have properties other than their quantum state. These properties (also called ``hidden variables") would provide  what we call a non-contextual value-map in subsection \ref{sec3.2}.
 
The main point of our paper is to show that no  choice between the rejection of ``realism" or of locality has to be made, since the existence of a non-contextual value-map is a {\it consequence} of locality and the perfect correlations. 

It is also true that such a  non-contextual value-map cannot exist (section \ref{sec4}). But, since the perfect correlations are a well-established empirical fact, the only logical conclusion is that locality is false.

So, ``realism" (in the sense of the existence of a non-contextual value-map) is not an assumption in our reasoning but is a consequence of the assumption of  locality (and the perfect correlations).

It is worth pointing out that this was also how Bell viewed his own discovery:\footnote{Here, ``determinism" refers to the idea of there being pre-existing 
 values $v(A)$ or of the existence of a non-contextual value-map.}

\begin{quote}

It is important to note that to the limited degree to which {\it determinism}
plays a role in the EPR argument, it is not assumed but {\it  inferred}.
What is held sacred is the principle of `local causality'---or `no action
at a distance'. [\dots] 

It is remarkably difficult to get this point across, that determinism
is not a {\it  presupposition} of the analysis.

\begin{flushright} John Bell \cite{BS} reprinted in  \cite[p. 143]{B} (italics in the original) \end{flushright}

\end{quote}
Bell was perfectly aware of  the misunderstandings of his theorem, so he added, unfortunately only in a footnote:
 
\begin{quote}

My own first paper on this subject ({\it Physics} {\bf 1}, 195 (1965))\footnote{Reference \cite{Be2}, reprinted as Chap. 2 in \cite{B}.} starts with a summary of the EPR argument {\it from locality to} deterministic hidden variables. But the commentators have almost universally reported that it begins with deterministic hidden variables.

\begin{flushright} John Bell \cite{BS} reprinted in \cite[p. 157, footnote 10]{B} (italics in the original) \end{flushright}

\end{quote}

This, however, did not stop the misunderstandings!

We mentioned in section \ref{sec1} some other works related to our paper; the relationship is that they also use a maximally entangled state and  a Kochen-Specker type argument (see Remark 2 in section \ref{sec4}) to derive a contradiction between locality and ``realism" and not simply  a ``non-existence of a non-contextual value-map" result as in Kochen and Specker \cite{KS}. This was remarked by Heywood and  Redhead \cite{H-R} in 1983, whose work was simplified and extended by Stairs \cite{St},  Brown and  Svetlichny \cite{Brown} and Elby \cite{El}. But, as far as we can see, those papers still conclude that their arguments refute ``local realism," namely they remain within the paradigm, ``reject realism or locality" (or both). 

On the other hand, both Cabello \cite{Cab} and Aravind \cite{Ara} establish a clear link between locality and the existence of a non-contextual value-map (in our language) and then use a variant of the Kochen-Specker argument to rule out the existence of such a map, hence to establish nonlocality.

There has also been quite a lot of  discussion about the ``free will theorem" of Conway and Kochen \cite{CK}, which is related to this paper in the sense that they also use a Kochen-Specker argument combined with the principle of perfect correlations to derive their conclusions. But what they claim to establish is that, if humans have ``free will," so do the particles, meaning that, if experimentalists are  free to choose which ``observable" to measure,\footnote{In fact, this is the property that we discussed  in remark 4 of subsection \ref{sec3.2}, and, as Schr\"odinger's example of the schoolchildren shows, it does not depend on humans having ``genuine free will" (whatever that means).} then no deterministic theory can account for the results, given some  assumptions that Conway and Kochen consider as very reasonable.\footnote{They also claim that their results rule out certain ``spontaneous collapse theories," a statement criticized in \cite{BG, Tu1, GTTZ}.}

Their assumptions are basically properties 1 and 3 in the Remark in section \ref{sec4}, perfect correlations and locality. These assumptions are respectively what
Conway and Kochen call SPIN, TWIN and FIN or MIN, depending on the version of their argument, see \cite{CK}. They show that those assumptions plus determinism and the ``free will" of the experimentalists lead to a  contradiction. But as pointed out (among other criticisms) in \cite{GTTZ} and in \cite[Section 4.5.2]{Hem-S},  properties 1 and 3 in the Remark in section \ref{sec4} (i.e. SPIN) and the perfect correlations (TWIN) suffice to prove non-locality (the negation of MIN), since a combination all three properties (SPIN, TWIN and MIN)  lead to a contradiction and SPIN and TWIN are empirical data. 

Moreover, as shown in this paper and in  \cite{Hem, Hem-S},
the  perfect correlations  (TWIN) alone establish non-locality\footnote{Since, combined with locality, they imply the existence of a non-contextual value-map, satisfying 
 (\ref{res}) and thus also (\ref{res2}, \ref{res3}), whose existence is impossible because of theorem \ref{2}.} so that the issue of determinism is simply not relevant here (moreover, as Bohmian mechanics shows, determinism is perfectly compatible with the quantum predictions).
 
To come back to the EPR dilemma of section \ref{sec2}  (do measurements reveal or create the reality indicated by their results?), the answer is that, in general, they do create that reality, in the sense described earlier,  but they do it in a nonlocal way.

An important lesson coming from our discussions here is that
 the concept of `measurement' as it appears in the traditional formulations of quantum
theory can often have a deleterious effect upon one's thinking and tends to cloud the issues. Here is John Stewart Bell on the topic:

\begin{quote}
I suspect \dots [physicists] \dots were misled by the use of the word `measurement' in contemporary theory. This 
word very strongly suggests the ascertaining of some preexisting property of some thing, any 
instrument involved playing a purely passive role. Quantum experiments are just not 
like that, as we learned especially from Bohr. The results have to be regarded as the joint-product of 
`system' and `apparatus,' the complete experimental setup.
 But the misuse of the word 
`measurement' make it easy to forget this and then expect that the `result of measurement' 
should obey some simple logic in which the apparatus is not mentioned. \dots I am convinced 
that the word `measurement' has now been so abused that the field would be significantly 
advanced by banning its use altogether, in favor for example of the word~`experiment.'

\begin{flushright} John Bell  \cite{Be7}
reprinted in  \cite[p. 166]{B}  \end{flushright}

\end{quote}

In Bohmian mechanics (section \ref{sec6}), one sees exactly what is going on: it is a realistic
 theory in the sense that particles have objective properties at
all times, not just when observed. The particles' description includes more than just their quantum state, namely also their positions. Yet, at the
same time, and in accord with Bell's suggestion, an
analysis of the experiments that are usually called measurements---but would better be called ``measurements"---of a particle or system shows that the latter are contextual:   they do not reveal a pre-existing property of the particle or system, but are genuine interactions between  system and `measuring' apparatus that  yield  results reflecting the state of the apparatus as well as that of the system. 

The contextuality of measurement would be absurd if the word measurement were taken in its usual sense, namely the discovery of the pre-existing value of the measured quantity.\footnote{As we mentioned in subsection \ref{sec3.2}, if there exists a   pre-existing value $v(A)$, then the result of any experiment  ${\cal E}_A$ measuring $A$ must be $v(A)$.} But the contextuality of quantum mechanical measurements, that do not reveal such values, so that the use of the word ``measurement" is a misuse and an abuse of that word, is almost a triviality.
However, when it occurs for well-separated systems in an entangled state, this contextuality is elevated to nonlocality, which is indeed a striking innovation, but one that is nonetheless an inevitable consequence of the quantum predictions.

 \end{document}